\documentclass[12pt]{article} 
\usepackage[top=30mm, bottom=40mm, left=25mm, right=25mm]{geometry}
\usepackage{cite} 

\usepackage{graphicx}
\usepackage{amsmath, amssymb} 				
\usepackage{empheq}
\usepackage{fancybox}
\usepackage{mathrsfs} 
\usepackage{textcomp} 
\usepackage{mathtools}

\usepackage{epstopdf}
\usepackage{bm} 
\usepackage{cases} 

\usepackage{latexsym}
\usepackage{marvosym}
\usepackage{bm}	

\usepackage[mathcal]{euscript}
\usepackage{indentfirst}

\usepackage[rgb]{xcolor}

\usepackage{hyperref}
\hypersetup{
	unicode,	
	colorlinks,
	citecolor=[rgb]{0,0,1},
	linkcolor=[rgb]{.5, 0, .5},
	urlcolor=[rgb]{.6,0,0},  
	bookmarksopen=true,
	bookmarksopenlevel=\maxdimen,
	bookmarksnumbered
	}

\numberwithin{equation}{section}

\usepackage[all]{xy}

\usepackage{amsthm} 
\theoremstyle{definition}

\theoremstyle{plain}

\font\biggest=cmssbx10 scaled 3150

\def\K3{\mathrm K3}

\def\gl#1#2{$\mathrm{GL}(#1; {\bf #2})$}
\def\sl#1#2{$\mathrm{SL}(#1; {\bf #2})$}
\def\sp#1#2{$\mathrm{Sp}(#1; {\bf #2})$}
\def\spin#1#2{$\mathrm{Spin}(#1, #2)$}
\def\su#1#2{SU({#1,#2})}
\def\usp#1#2{USp({#1,#2})}

\def\double #1{#1{\hbox{\kern-2pt $#1$}}}
\def\un#1{\underline #1}

\def\gl#1#2{\ifmmode \mathrm{GL}(#1; {\bf #2}) \else $\mathrm{GL}(#1; {\bf #2})$\fi}
\def\sl#1#2{\ifmmode \mathrm{SL}(#1; {\bf #2}) \else $\mathrm{SL}(#1; {\bf #2})$\fi}
\def\so#1{\ifmmode \mathrm{SO}({#1}) \else $\mathrm{SO}(#1)$\fi}

\def\sp#1#2{\ifmmode \mathrm{Sp}(#1; {\bf #2}) \else $\mathrm{Sp}(#1; {\bf #2})$\fi}
\def\usp#1#2{\ifmmode \mathrm{USp}(#1,#2) \else $\mathrm{USp}(#1,#2)$\fi}
\def\spin#1#2{\ifmmode \mathrm{Spin}(#1,#2) \else $\mathrm{Spin}(#1,#2)$\fi} 
\def\su#1{\ifmmode \mathrm{SU}({#1}) \else $\mathrm{SU}(#1)$\fi}

\def\on#1#2{{\buildrel{{\mkern2.5mu\raise-.1em\hbox{$\scriptstyle#1$}\mkern-2.5mu}}\over{#2}}}		
\def\ron#1#2{{\buildrel{{\raise-.1em\hbox{$\scriptstyle#1$}}}\over{#2}}}		
\def\dt#1{\on{\hbox{\bf .}}{#1}}                
\mathcode`\*="702A                  
\def\f#1#2{{\textstyle{#1\over#2}}}	   
\def\half{{\textstyle{1\over{\raise.1ex\hbox{$\scriptstyle{2}$}}}}}
\def\slap#1#2{\setbox0=\hbox{$#1{#2}$}#2\kern-\wd0{\hbox to\wd0{\hfil$#1{/}$\hfil}}}
\def\sla#1{\mathpalette\slap{#1}}		
\def\bop#1{\setbox0=\hbox{$#1M$}\mkern1.5mu
	\vbox{\hrule height0pt depth.1\ht0
	\hbox{\vrule width.1\ht0 height.8\ht0 \kern.8\ht0
	\vrule width.1\ht0}\hrule height.1\ht0}\mkern1.5mu}
\def\bo{{\mathpalette\bop{}}}                        

\def\Gamma{\mathchar"0100}
\def\Delta{\mathchar"0101}
\def\Theta{\mathchar"0102}
\def\Lambda{\mathchar"0103}
\def\Xi{\mathchar"0104}
\def\Pi{\mathchar"0105}
\def\Sigma{\mathchar"0106}
\def\Upsilon{\mathchar"0107}
\def\Phi{\mathchar"0108}
\def\Psi{\mathchar"0109}
\def\Omega{\mathchar"010A}

\catcode128=13 \def €{\"A}                 
\catcode129=13 \def {\AA}                 
\catcode130=13 \def '{\c}           	   
\catcode131=13 \def ƒ{\'E}                   
\catcode132=13 \def "{\~N}                   
\catcode133=13 \def …{\"O}                 
\catcode134=13 \def †{\"U}                  
\catcode135=13 \def ‡{\'a}                  
\catcode136=13 \def ˆ{\`a}                   
\catcode137=13 \def ‰{\^a}                 
\catcode138=13 \def Š{\"a}                 
\catcode139=13 \def ‹{\~a}                   
\catcode140=13 \def Œ{\alpha}            
\catcode141=13 \def {\chi}                
\catcode142=13 \def Ž{\'e}                   
\catcode143=13 \def {\`e}                    
\catcode144=13 \def {\^e}                  
\catcode145=13 \def '{\"e}                
\catcode146=13 \def '{\'\i}                 
\catcode147=13 \def "{\`\i}                  
\catcode148=13 \def "{\^\i}                
\catcode149=13 \def •{\"\i}                
\catcode150=13 \def –{\~n}                  
\catcode151=13 \def —{\'o}                 
\catcode152=13 \def ˜{\`o}                  
\catcode153=13 \def ™{\^o}                
\catcode154=13 \def š{\"o}                 
\catcode155=13 \def ›{\~o}                  
\catcode156=13 \def œ{\'u}                  
\catcode157=13 \def {\`u}                  
\catcode158=13 \def ž{\^u}                
\catcode159=13 \def Ÿ{\"u}                
\catcode160=13 \def  {\tau}               
\catcode162=13 \def ¢{\oplus}           
\catcode163=13 \def £{\relax\ifmmode\expandafter\to\else\expandafter\itemize\fi} 
\catcode164=13 \def ¤{\subset}	  
\catcode165=13 \def ¥{\infty}           
\catcode166=13 \def ¦{\mp}                
\catcode167=13 \def §{\sigma}           
\catcode168=13 \def ¨{\rho}               
\catcode169=13 \def ©{\gamma}         
\catcode170=13 \def ª{\leftrightarrow} 
\catcode171=13 \def «{\relax\ifmmode\expandafter\acute\else\expandafter\'\fi}
\catcode172=13 \def ¬{\relax\ifmmode\expandafter\ddt\else\expandafter\"\fi}
\catcode173=13 \def ­{\equiv}            
\catcode174=13 \def ®{\approx}          
\catcode175=13 \def ¯{\Omega}          
\catcode176=13 \def °{\otimes}          
\catcode177=13 \def ±{\ne}                 
\catcode178=13 \def ²{\le}                   
\catcode179=13 \def ³{\ge}                  
\catcode180=13 \def ´{\upsilon}          
\catcode181=13 \def µ{\mu}                
\catcode182=13 \def ¶{\delta}             
\catcode183=13 \def ·{\epsilon}          
\catcode184=13 \def ¸{\Pi}                  
\catcode185=13 \def ¹{\pi}                  
\catcode186=13 \def º{\beta}               
\catcode187=13 \def »{\partial}           
\catcode188=13 \def ¼{\nobreak\ }       
\catcode189=13 \def ½{\zeta}               
\catcode190=13 \def ¾{\sim}                 
\catcode191=13 \def ¿{\omega}           
\catcode192=13 \def À{\dt}                     
\catcode193=13 \def Á{\gets}                
\catcode194=13 \def Â{\lambda}           
\catcode195=13 \def Ã{\nu}                   
\catcode196=13 \def Ä{\phi}                  
\catcode197=13 \def Å{\xi}                     
\catcode198=13 \def Æ{\psi}                  
\catcode199=13 \def Ç{\int}                    
\catcode200=13 \def È{\oint}                 
\catcode201=13 \def É{\relax\ifmmode\expandafter\cdot\else\vol\fi}    
\catcode203=13 \def Ë{\`A}                      
\catcode204=13 \def Ì{\~A}                      
\catcode205=13 \def Í{\~O}                      
\catcode206=13 \def Î{\Theta}              
\catcode207=13 \def Ï{\theta}               
\catcode208=13 \def Ð{\relax\ifmmode\expandafter\bar\else\expandafter\=\fi}
\catcode209=13 \def Ñ{\overline}             
\catcode210=13 \def Ò{\langle}               
\catcode211=13 \def Ó{\relax\ifmmode\expandafter\{\else\ital\fi}      
\catcode212=13 \def Ô{\rangle}               
\catcode213=13 \def Õ{\}}                        
\catcode214=13 \def Ö{\sla}                      
\catcode215=13 \def ×{\relax\ifmmode\expandafter\check\else\expandafter\v\fi}
\catcode216=13 \def Ø{\"y}                     
\catcode217=13 \def Ù{\"Y}  		    
\catcode218=13 \def Ú{\Leftarrow}       
\catcode219=13 \def Û{\Leftrightarrow}       
\catcode220=13 \def Ü{\relax\ifmmode\expandafter\Rightarrow\else\sect\fi}
\catcode221=13 \def Ý{\sum}                  
\catcode222=13 \def Þ{\prod}                 
\catcode223=13 \def ß{\widehat}              
\catcode224=13 \def à{\pm}                     
\catcode225=13 \def á{\nabla}                
\catcode226=13 \def â{\quad}                 
\catcode227=13 \def ã{\in}               	
\catcode228=13 \def ä{\star}      	      
\catcode229=13 \def å{\sqrt}                   
\catcode230=13 \def æ{\^E}			
\catcode231=13 \def ç{\Upsilon}              
\catcode232=13 \def è{\"E}    	   	 
\catcode233=13 \def é{\`E}               	  
\catcode234=13 \def ê{\Sigma}                
\catcode235=13 \def ë{\Delta}                 
\catcode236=13 \def ì{\Phi}                     
\catcode237=13 \def í{\`I}        		   
\catcode238=13 \def î{\iota}        	     
\catcode239=13 \def ï{\Psi}                     
\catcode240=13 \def ð{\times}                  
\catcode241=13 \def ñ{\Lambda}             
\catcode242=13 \def ò{\cdots}                
\catcode243=13 \def ó{\^U}			
\catcode244=13 \def ô{\`U}    	              
\catcode245=13 \def õ{\bo}                       
\catcode246=13 \def ö{\relax\ifmmode\expandafter\hat\else\expandafter\^\fi}
\catcode247=13 \def÷{\relax\ifmmode\expandafter\tilde\else\expandafter\~\fi}
\catcode248=13 \def ø{\ll}                         
\catcode249=13 \def ù{\gg}                       
\catcode250=13 \def ú{\eta}                      
\catcode251=13 \def û{\kappa}                  
\catcode252=13 \def ü{\half}     		 
\catcode253=13 \def ý{\Gamma} 		
\catcode254=13 \def þ{\Xi}   			
\catcode255=13 \def ÿ{\relax\ifmmode\expandafter{}^{\dagger}{}\else\dag\fi}

\def\A{{\cal A}}  \def\B{{\cal B}}  \def\C{{\cal C}}  \def\D{{\cal D}}
      \def\H{{\cal H}}   
  \def\K{{\cal K}}  \def\L{{\cal L}}       
      
\def\S{{\cal S}}    \def\U{{\cal U}} \def\V{{\cal V}}

\font\textscr=stix-mathscr at 12pt \font\scriptscr=stix-mathscr at 8pt \font\sscriptscr=stix-mathscr at 6pt
\def\scr{\fam15 \textscr}
		
\let\ced=\c		\def '{\ced}		\def\c{{\scr c}}	
	
	\def\ff{{\scr f}}

\def\^{\wedge}
\def\mapright#1{\smash{
            \mathop{\longrightarrow}\limits^{#1}}}	

\def\dig#1{\setbox0=\hbox{$#1M$}
	\hskip.06\wd0 \vrule width.07\wd0 height.63\wd0 depth.01\wd0 
	\vrule width.37\wd0 height.63\wd0 depth-.56\wd0 \hskip-.4\wd0
	\vrule width.25\wd0 height.35\wd0 depth-.28\wd0 
	\vrule width.07\wd0 height.35\wd0 depth-.17\wd0 \hskip.14\wd0}
\def\digamma{{\mathpalette\dig{}}}

\def\hang{\hangindent\parindent}
\def\bitem{\par\hang\textindent}

\def\textindent#1{\indent\llap{#1\enspace}\ignorespaces}

\newcount\hrs
 \newcount\mins
 \newcount\merid
 \newcount\hrmins
 \newcount\hrmerid
 \hrs=\time
 \mins=\time
 \divide\hrs by 60
 \merid=\hrs
 \hrmins=\hrs
 \divide\merid by 12
 \hrmerid=\merid
 \multiply\hrmerid by 12
 \advance\hrs by -\hrmerid
 \ifnum\hrs=0\hrs=12\fi
 \multiply\hrmins by 60
 \advance\mins by -\hrmins
 
\def\watch{\number\hrs:\ifnum\mins<10 {0}\fi\number\mins\space\ifnum\merid=0 AM\else PM\fi}

{\catcode`\D=12
 \gdef\getYear D:#1#2#3#4{\edef\xYear{#1#2#3#4}\getMonth}
}
\def\getMonth#1#2{\edef\xMonth{#1#2}\getDay}
\def\getDay#1#2{\edef\xDay{#1#2}\getHour}
\def\getHour#1#2{\edef\xHour{#1#2}\getMin}
\def\getMin#1#2{\edef\xMin{#1#2}\getSec}
\def\getSec#1#2{\edef\xSec{#1#2}\getTZh}
\def\getTZh #1#2#3{\edef\xSign{#1}\edef\xTZh{#2#3}\getTZm}
\def\getTZm '#1#2'{%
    \edef\xTZm{#1#2}%
    \edef\convDate{\xYear-\xMonth-\xDay T\xHour:\xMin:\xSec-\xTZh:\xTZm}
    \edef\dated{\xHour:\xMin:\xSec \ GMT\xSign\xTZh:\xTZm, \xMonth/\xDay/\xYear}
    \edef\zone{$\xSign$\xTZh:\xTZm}%
}
 
\catcode`\|=\active
\def|{\relax\ifmmode\delimiter"026A30C \else$\mathchar"026A$\fi}

\title{
{\color[rgb]{.5,0,1}\biggest Enlarged exceptional symmetries \\[.1in]
of first-quantized F-theory}
}
\author{\href{mailto:siegel@insti.physics.sunysb.edu}{Warren Siegel} and \href{mailto:di.wang.2@stonybrook.edu}{Di Wang}}
\date{June 6, 2018
}		

\begin{document}
\maketitle

\vspace*{-9cm}
\begin{flushright}
{~\\
YITP-SB-18-15}
\end{flushright}
\vspace*{+7cm}

\begin{center}
\vskip -.5in
{\em
C. N. Yang Institute for Theoretical Physics\\
State University of New York, Stony Brook, NY 11794-3840
}
\end{center} 

\vspace{5pt}

\begin{abstract}
The exceptional symmetries of supergravity have been reproduced from the Hamiltonian formulation of the classical mechanics of F-theory.
We now find the Lagrangian formalism has even larger exceptional symmetries, simplifying its derivation:
We discuss D = 5 as an example.
\end{abstract}

\vfill

\setcounter{page}1
\thispagestyle{empty}
\newpage

{
\tableofcontents
}

\newpage

\section{Introduction}
\label{S:Introduction}

F-theory is a proposal to make STU-duality (at least) as manifest on the entire string (not just supergravity) as T-duality is in ordinary string theory (i.e., visible in the first-quantized action).
F-theory, M(embrane)-theory, T(-duality)-theory \cite{Siegel:1993th,Siegel:1993bj}, and S(tring)-theory are related \cite{Polacek:2014cva}
by spontaneous symmetry breaking, reducing the dimensions of both spacetime (D) and the worldvolume (d), and the manifest symmetry, in ``unitary" gauges for the ``section" conditions.
(The section conditions that reduce F-theory to M-, T-, or S-theory result from replacing some of the currents in the constraints with their zero-modes, acting on functions or their products.)
Dualities can mix the eliminated worldvolume dimensions with retained ones, hence exchanging strings with branes.
The worldvolume ``field strengths" of the spacetime coordinates of F-theory satisfy a selfduality condition \cite{Linch:2015fya, Linch:2015qva,Linch:2015fca};
the resulting version of S-theory has double the spacetime coordinates, each satisfying worldsheet (anti)selfduality \cite{Duff:1989tf,Tseytlin:1990nb,Tseytlin:1990va}.
(Normal S-theory, and the original version of T-theory, have the usual number D of spacetime coordinates, but compactification produces D + D zero-modes: D momenta and D winding.  They also have D + D currents, from $§$ and $ $.)

Dualities are both spacetime and Weyl reflections for the corresonding ``spin group".
For S-theory the group is simply the Lorentz group O(D$-$1,1), and a close relative for M- (O(D,1)) and T-theory (O(D$-$1,1)$^2$); for F-theory it is various generalizations, related to doubling the size of the covering group of the Lorentz group.
The F-symmetry is expected to persevere in ``covariant" gauges, generalizing the conformal gauge in S-theory, where section conditions are applied only on external states via Becchi-Rouet-Stora-Tyutin \cite{Siegel:2016dek}.

One interesting way in which F-theory differs from standard S-theory is that the Lorentz symmetry of spacetime acts directly on the worldvolume coordinates, not just the spacetime coordinates.  This is an analog of extended worldsheet supersymmetry in the Ramond-Neveu-Schwarz formalism for superstrings \cite{Ademollo:1976pp} (describing maximally supersymmetric spacetime theories that are selfdual and have critical dimension D = 4 \cite{Ooguri:1990ww,Ooguri:1991fp,Marcus:1992xt,Siegel:1992ev}), where the R-symmetry of the spacetime supersymmetry algebra is equated with the R-symmetry of the worldsheet superVirasoro algebra.
(In the latter theories the corresponding R-symmetry currents are quadratic in the fermionic worldvolume ``fields".)

In this paper we solve a problem encountered in earlier treatments of the worldvolume Lagrangian ($\L$) formalism of F-theory:
We find a larger Lagrangian symmetry of the worldvolume theory (``F-symmetry") which includes as subgroups not only the ``Lorentz" symmetry of the Lagrangian, but also the exceptional group symmetry of the Hamiltonian ($\H$) formalism.  This new grouping of exceptional groups not only better unifies the $\L$ and $\H$ formalisms, but allows us to give for the first time a manifestly covariant formulation of the D = 5 $\L$ formalism.


\section{Exceptional Lagrangian symmetries}


\subsection{F-symmetry}

When constructing a fully covariant $\L$ formalism, it will prove economical to consider an ``off-shell" group ``F" containing as (the previously described) subgroups both the symmetry ``L" of the Lagrangian and the exceptional symmetry ``G" of the bosonic spacetime coordinates of the Hamiltonian.
L and ``H" are Wick rotations of the maximal compact subgroups of F and G, up to Abelian factors.  (E.g., SL(n) $\supset$ SO(n), SO(n,n) $\supset$ SO(n)$^2$.  These are also the cosets used for gravity and T-theory, respectively.)
$$
\vcenter{\halign{ # && # \cr
	& \multispan3exceptionalâ & tangent \cr
	$\L$(agrangian) &â& F & \hfil $\supset$ \hfil & L \cr
	&& $\cup$ && $\cup$ \cr
	$\H$(amiltonian) && G & \hfil $\supset$ \hfil & H \cr
	}}
âââ
\vcenter{\halign{ # & #\hfil \cr
	$\cup$ & $d\supset d-1$ \cr
	$\supset$ & maximal compact subgroup \cr
	}}
$$
The exceptional symmetries F and G act only on the bosons.
The ``tangent-space" symmetries L of the Lagrangian and H of the Hamiltonian apply to the fermions, and to the bosons as subgroups of F and G.
(The classical group H is related to the covering group of the spacetime Lorentz group with doubled argument.
L follows from it by dropping the Sp/SO constraint if there, doubling the group otherwise.)
The coset G/H describes a generalization of gravity for the background fields.

The F group's Dynkin diagram is the G group's with an extra node on the ``short" leg.
Thus increasing the rank by 1 adds $ $ to the worldvolume coordinates.
(Underlining indicates the growth from $\H¤\L$; thus $\un §=( ,§)$.)
\begin{subequations}
\begin{align}
\label{Dynk}
\hskip-2in \rm G\hskip2in
\resizebox{2mm}{!}{
$\thicklines
\put(0,0){\circle{10}}
\put(5,0){\line(1,0){35}}
\put(45,0){\circle{10}}
\put(50,0){\line(1,0){35}}
\put(90,0){\circle{10}}
\put(0,5){\line(0,1){35}}
\put(0,45){\circle{10}}
\put(-25,0){\line(1,0){20}}
\put(-50,0){\circle*{2}}\put(-45,0){\circle*{2}}\put(-40,0){\circle*{2}} 
\put(-85,0){\line(1,0){20}}
\put(-90,0){\circle{10}}
\put(-118,-3){\resizebox{5.5mm}{!}{{$X$}}}
\put(105,-5){\resizebox{3.5mm}{!}{{$§$}}}
\put(15,40){\resizebox{3.5mm}{!}{{$Â$}}}
%
$
} 
\\[.2in]
\label{Dynki}
\hskip-2in \rm F\hskip2in
\resizebox{2mm}{!}{
$\thicklines
\put(0,0){\circle{10}}
\put(5,0){\line(1,0){35}}
\put(45,0){\circle{10}}
\put(50,0){\line(1,0){35}}
\put(90,0){\circle{10}}
\put(95,0){\line(1,0){35}}
\put(135,0){\circle{10}}
\put(0,5){\line(0,1){35}}
\put(0,45){\circle{10}}
\put(-25,0){\line(1,0){20}}
\put(-50,0){\circle*{2}}\put(-45,0){\circle*{2}}\put(-40,0){\circle*{2}} 
\put(-85,0){\line(1,0){20}}
\put(-90,0){\circle{10}}
\put(-118,-3){\resizebox{5.5mm}{!}{{$\un X$}}}
\put(150,-5){\resizebox{14mm}{!}{{$§¢ $}}}
\put(15,40){\resizebox{3.5mm}{!}{{$\un Â$}}}
%
$
} 
\end{align}
\end{subequations}
(Equivalently, by flipping the E$_7$ Dynkin diagram, the series for F and G correspond to changing its length on different legs.)
Thus F$_{\rm D}\supset$ F$_{{\rm D}-1}$ by killing a node on one end, but $\supset$ G$_{\rm D}$ on the other.
In this notation, the rank of G$_{\rm D}=$ E$_{\rm D+1}$ is D+1, while the rank of F$_{\rm D}$ is D+2.
This is the sense is which S-theory in D dimensions corresponds to M-theory in D+1 dimensions and F-theory in D+2 dimensions.

The field equations of the bosons are selfduality equations, a generalization of those of (anti)chiral bosons of the worldsheet, leading to the F-theory generalization of the T-duality of T-theory.
Thus all bosons appearing in the first-quantized Lagrangian are representations of F, but the (free) selfduality equations (or vacuum) spontaneously break this symmetry to L.
The situation is similar in the $\H$ formalism, where the kinematics are G symmetric, but the (free) dynamics are only H symmetric.
In all the above cases, the background fields restore the full symmetry.
Thus the field space of the background is the coset, and couples only to the dynamics.

A relevant analog of these classical mechanics (or first-quantization) groups in classical field theory (or second-quantization) is the vector fields of 4D N = 8 supergravity.  There the field equations can be expressed as selfduality of a formulation with both electric (polar) and magnetic (axial) 4-vector potentials.  Selfduality can be written entirely with ``curved" GL(4) (``$m$") and E$_7$ (``$m'$", {\bf 56}) indices as
\begin{equation}
(å{-g}g^{mp}g^{nq})g^{m'n'}F_{pqn'} ­ ÷F^{mnm'} = ü·^{mnpq}C^{m'n'}F_{pqn'}
\end{equation}
where $C^{m'n'}$ is the E$_7$-invariant Sp(56) metric and $F_{mnm'}$ is the ordinary curl $»_{[m}A_{n]m'}$.  (The non-selfdual Lagrangian would be $¾F÷F$.)  The spontaneous breaking of GL(4) $£$ SO(3,1) and E$_7£$ SU(8) comes from the vacuum values of the 2 types of symmetric metric $g_{mn}$ (gravity) and $g_{m'n'}$ (scalars), or from the fact they are required to be group elements.  Alternatively, we can manifest the local SO(3,1) and SU(8) by using ``flat" indices:
\begin{equation}
g_{mn} = e_m{}^a e_n{}^b ú_{ab}¼,âg_{m'n'} = e_{m'}{}^{a'} e_{n'}{}^{b'} ú_{a'b'}
\end{equation}
where the 2 types of vielbein are elements of the cosets GL(4)/SO(3,1) ($e_a{}^m$) and E$_7$/SU(8) ($e_{a'}{}^{m'}$), and the flat metrics $ú$ are symmetric, invariant tensors of SO(3,1) and SU(8) (for the {\bf 56} $£$ {\bf 28} + $Ñ{\bf 28}$).
Selfduality can then be expressed as
\begin{equation}
e_a{}^m e_b{}^n e_{a'}{}^{m'} F_{mnm'} ­ F_{aba'} = (ú_{ac}ú_{bd}ú_{a'b'})(ü·^{cdef}C^{b'c'})F_{efc'}
\end{equation}
Thus there are 56 independent vectors off shell, but only 28 on, corresponding to the fact that there are 28 states with helicity +1 and 28 with $-$1.  This is analogous to the doubled-dimension version of S-theory, where there are D + D spacetime coordinates off shell but D on, describing D left-handed sets of modes and D right.
(Similar constructions apply for differential forms of other ranks and in other dimensions.)


\subsection{Cases}

The well-understood cases of these symmetries are:
$$
\vcenter{\halign{\hfil#&\hfil#\hfil&#\hfil\cr
	& F &\cr
	\raise.5em\hbox{\color{blue}$ $} $\swarrow$ & & $\searrow$ \raise.5em\hbox{\color{blue}$|0Ô$}\cr
	Gâ & & âL \cr
	$\searrow$¼& & $\swarrow$ \cr
	& H &\cr}}
âââ
\vcenter{
\halign{ # && â#\hfil \cr
D & d & F & G & L & H \cr
0 & 2 & GL(2) & GL(1) & GL(1,{\bf C}) & I \cr
1 & 3 & GL(3) & GL(2) & GL(2) & SO(1,1) \cr
2 & 4 & SL(4)SL(2) & SL(3)SL(2) & GL(2)$^2$ & GL(2) \cr
3 & 6 & SL(6) & SL(5) & GL(4) & Sp(4) \cr
4 & 12 & SO(6,6) & SO(5,5) & GL(4,{\bf C}) & Sp(4,{\bf C}) \cr
5 & 56 & E$_{7(7)}$ & E$_{6(6)}$ & U*(8) & USp(4,4) \cr
6 & ? & ? & E$_{7(7)}$ & U*(8)$^2$ & SU*(8) \cr
7 & ? & ? & E$_{8(8)}$ & U*(16) & SO*(16) \cr
}
} $$
where D is the number of spacetime dimensions, while d is the number of worldvolume dimensions, including $ $.
(The ?'s may be infinite dimensional.
``$ $" refers to breaking by going from $\L£\H$ formalism.
``$|0Ô$" is breaking by the vacuum value of the background in the equations of motion, as generated by selfduality in the $\L$ formalism or the Hamiltonian in its formalism.)
Note that G$ð$GL(1) $\subset$ F, where GL(1) is related to $ $.
Also, L always includes a GL(1).

The dimensions of these groups satisfy 
\begin{subequations}
\begin{align}
[{\rm G}]-2[{\rm H}] & = {\rm D}+1 \\
[{\rm F}]-2[{\rm L}] & = {\rm D} \\
\label{E:selfd}
[{\rm L}]-[{\rm H}] & = {\rm d}
\end{align}
\end{subequations}
(For the last, we need to impose selfduality of the worldvolume for D = 5, so 56 $£$ 28:  See below.)
Thus
$$
\vcenter{\halign{#&&#\cr
	& \multispan2 \hfill [F] = 2[H] + D + 2d & \cr
	& \hfil $\swarrow$âââ & $\searrow$â¼ & \cr
	\multispan2 [G] = 2[H] + D + 1ââââ¼ & \multispan2 â[L] = [H] + d \cr
	& \hfil $\searrow$âââ & $\swarrow$â¼ & \cr
	& \multispan2 ââââ¼[H] &\cr}}
$$
This leads to identities such as
\begin{subequations}
\begin{align}
\label{E:adj}
[{\rm F}] & = [{\rm G}ð{\rm GL(1)}] + 2({\rm d} - 1) \\
[{\rm L/H}] & = {\rm d} \\
[{\rm F/L}] & = [{\rm G/H}] + ({\rm d} -1)
\end{align}
\end{subequations}

These F groups are the same, up to Wick rotation, as the exceptional groups for 4D N-extended supergravity for various N, where N = 2 + the D of F (and N = 7 is equivalent to N = 8):
$$
\vcenter{
\halign{ # && â#\hfil \cr
D & F$_{\rm D}$ & E$_{\rm N}$ & N \cr
0 & GL(2) & U(2) & 2 \cr
1 & GL(3) & U(3) & 3 \cr
2 & SL(4)SL(2) & SU(4)SU(1,1) & 4 \cr
3 & SL(6) & SU(5,1) & 5 \cr
4 & SO(6,6) & SO*(12) & 6 \cr
5 & E$_{7(7)}$ & E$_{7(7)}$ & 7 (8) \cr
}
} $$
\vskip.1in
\noindent This is apparently related to the ``disintegration triangle" \cite{Julia:1980gr}.


\subsection{Representations}

We find the representations of important quantities (mostly) from (\ref{Dynki}): worldvolume derivatives $»$ (of number d), spacetime coordinates $X$, their gauge parameter $Â$ and field strength $F$, and worldvolume sectioning $\V$,
\begin{equation}
\label{4ms}
\vcenter{
\halign{ # & â# & â# && â$#$\hfil \cr
D & F & pattern & » & Â\mapright» & X\mapright» & F & \V \cr
0 & GL(2) & forms & 2 & 0 & 1 & 2 & \cr
1 & GL(3) & forms & 3 & 0¢1 & 1¢3 & 3¢3' & \cr
2 & SL(4)SL(2) & forms & (4,1) & (1,2) & (4,2) & (6,2) & \cr
3 & SL(6) & forms & 6 & 6 & 15 & 20 & \cr
4 & SO(6,6) & spinors & 12 & 32 & 32' & 32 & 1 \cr
5 & E$_{7(7)}$ & infinite & 56 & 912 & 133 & 56 & 133 \cr
}
} 
\end{equation}
(``0" $Â$ is a constant, for a global symmetry of scalars.)

Here the chains of representations take the form
\begin{equation}
»: ... £ Â £ X £ F £ B £ BB £ ...
\end{equation}
where the field strengths $F$ are (anti)selfdual representations of F, while the Bianchi identities $B$ are dual to the spacetime coordinates $X$, the Bianchi identities of the Bianchi identities $BB$ are dual to the gauge parameters $Â$, and the series may continue if the gauge invariances have their own gauge invariances.
Thus for D = 3 the series begins with a singlet (gauge invariance)$^2$ (as for T-theory) and terminates with the dual 6-form, while those for D = 4 repeat indefinitely in both directions, and D = 5 grows symmetrically away from the minimum for the field strength $F$ (e.g., the gauge invariance for the gauge invariance is {\bf 133 $¢$ 8645}
= ({\bf 133 $°$ 133})$_A$).

However, the field equations break the symmetry F $£$ L (as well as breaking G $£$ H):  The F-selfdual representations $F$ need to be separated into the L-selfdual and L-antiselfdual representations $F^{(à)}$.
For higher dimensions the worldvolume coordinates initially will receive a doubling for the same reason, since their naive L-representations are already chiral.

The branching for this symmetry breaking F $£$ L is:
\begin{equation}
\label{FtoL}
\vcenter{
\halign{ # & â# && â$#$\hfil \cr
D & L & » & Â & X & F \cr
0 & GL(1,{\bf C}) & 1¢Ð1 & 0 & 1 & 1¢Ð1 \cr
1 & GL(2) & 3 & 0¢1 & 1¢3 & 3¢3 \cr
2 & GL(2)$^2$ & (2,2) & 2(1) & 2(2,2) & 2(1,3)¢2(3,1) \cr
3 & GL(4) & 6 & 6 & 15 & 10¢10' \cr
4 & GL(4,{\bf C}) & 6¢Ð6 & 16¢16' & 16_C ¢ Ñ{16}_C & 16¢16' \cr
5 & U*(8) & 28¢28' & 36¢36'¢420¢420' & 63¢70 & 28¢28' \cr
}
} 
\end{equation}

On the other hand, reduction to the $\H$ formalism separates out $ $, breaking F¼$£$¼G (as well as breaking L $£$ H):

$$ \vcenter{
\halign{ # & ¼#\hfil && â$#$\hfil \cr
D & G & » & Â & X_  & X_§ & F \cr
0 & GL(1) & 1¢1 & 0 & 0 & 1 & 1¢1 \cr
1 & GL(2) & 1¢2 & 1 & 1 & 1¢2 & 2(1)¢2(2) \cr
2 & SL(3)SL(2) & 1¢(3,1) & (1,2) & (1,2) & (3,2) & (3,2)¢(3',2) \cr
3 & SL(5) & 1¢5 & 1¢5 & 5 & 10 & 10¢10' \cr
4 & SO(5,5) & 2(1)¢10 & 16¢16' & 16 & 16' & 16¢16' \cr
5 & E$_{6(6)}$ & 2(1)¢27¢27' & 27¢27'¢2(78) & 27'¢78 & 1¢27 & 2(1)¢27¢27' \cr
 & & & \hfill ¢351¢351' & & \cr
}
} $$
where for $X$ we have separated the ``Lagrange multiplier" $X_ $, corresponding to the primary gauge parameters (compare the $Â$ column) and dual to the Gauss constraints $\U$, from the ``physical" $X_§$, corresponding to the dual of the selfdual half of the field strengths (compare the $F$ column).

The previous treatment \cite{Linch:2016ipx} was incomplete:  Gauge invariances of the gauge invariances were ignored.  These are necessary for a covariant $\L$ formalism.  They already appear in T-theory for spacetime gauge transformations as a result of $\S$ sectioning.  This is a consequence of the fact that the 2-form time components do not have simply $¶B_{0i}=ÀÂ_i$, but also a term $-»_i Â_0$ for Lorentz covariance, which introduces a new gauge invariance of the gauge invariance $¶Â_a=»_aöÂ$, where $öÂ$ effectively cancels $Â_0$.  Thus $Â$ has additional components not seen from $X_ $. 

The splitting of $F$ under L and G isn't the same:  Under G it's into $ $ and $§$ pieces, but under L it's into selfdual and antiselfdual pieces, which are the sum and difference of the $ $ and $§$ pieces:
\begin{equation}
\label{E:pm}
F^{(à)} = ú^0 F_  à F_§ £ P à F_§
\end{equation}
($ú^0$ is a metric that breaks G $£$ H; but it's absorbed into the definition of the momenta $P$ conjugate to $X_§$ in the $\H$ formalism, restoring G symmetry.)


\section{D \texorpdfstring{$²$}{\textleq} 4}

We first translate previous results for lower dimensions D = 0-4 \cite{Linch:2015fya, Linch:2015qva,Linch:2015fca}, which were already in L-covariant form, into F-covariant form.
This does not require the introduction of any new representations nor the enlargement of any old ones, only the recognition that all the L representations  used previously can be combined into F ones.
We have there the approximate correspondence (ignoring Abelian factors, and the ``internal" SL(2) for D = 2)
\begin{equation}
\label{0th}
\vcenter{
\halign{ # && â#\hfil \cr
D & F & G & L & H \cr
0-3 & GL(d) & GL(d$-$1) & SO(d) & SO(d$-$1) \cr
4 & SO($\f{\rm d}2$,$\f{\rm d}2$) & SO($\f{\rm d}2-$1,$\f{\rm d}2-$1) & SO($\f{\rm d}2$,{\bf C}) & SO($\f{\rm d}2-$1,{\bf C}) \cr
}
}
\end{equation}
which makes the cases D = 0-3 resemble gravity for d dimensions, and D = 4 resemble T-theory for d/2 dimensions, (i.e., the worldvolume and not spacetime) for F/L, and decreasing the dimension by 1 for G/H (since the $\H$ theory lacks $ $).

The cases D = 0-3 can be transcribed immediately, since the enlargement of symmetry from L $£$ F is the usual for differential forms (\ref{0th},\ref{4ms}):  F has only the Levi-Civita tensor, while L has also the flat metric.

The case D = 4 doesn't require much more work:
It's the obvious generalization of the way in the $\H$ formalism matrices of H = Sp(4,{\bf C}) were combined into spinors of G = SO(5,5) in \cite{Linch:2015qva}.
Now the L = SL(4,{\bf C}) representations are all 4 $ð$ 4 bispinor matrices of various realities and symmetries \cite{Linch:2015qva},
and combine into F = SO(6,6) representations as in (\ref{FtoL}) vs.¼(\ref{4ms}).
Again $»$ acts on the chain of representations by hitting a Weyl spinor with $Ö»$, flipping the chirality, while the $\V$ constraint appears as $Ö»^2¾ õ$.


\section{D = 5}


\subsection{Lagrangian approach}

The use of F-symmetry allows us to easily complete the $\L$ formalism for D = 5, which in \cite{Linch:2016ipx} was described only in G-covariant language.
From (\ref{Dynki}) we identify $F_\A$ as the same representation {\bf 56} of F-symmetry E$_7$ as $»_\A$, while $X^A$ is in the adjoint {\bf 133}, and $Â_{\rm A}$ is a {\bf 912}.  
(The latter two index conventions will apply for just this subsection.)
The form of the gauge transformations and field strength then follow from just inserting the appropriate Clebsch-Gordan-Wigner coefficients:
\begin{align}
\label{E:para}
¶X_A & = º_A{}^{\B\rm C}»_\B Â_{\rm C} \\
F_\A & = \ff^{B\C}{}_\A »_\C X_B
\end{align}
($\ff$ is also the {\bf 56} matrix representation of the generators of the Lie algebra, with adjoint index raised by the Sp(56) metric.)
Gauge invariance of $F$ follows from the $\V$ constraint and algebraic identity for this case 
\begin{align}
\V_A ­ ü\ff_A{}^{\B\C}»_\B »_\C & = 0\\
\label{E:beta}
\ff^{D(\A\B} º_D{}^{\C)\rm E} & = 0
\end{align}
(The latter equation can be proven, e.g., by breaking E$_7£$U*(8):  See Appendix \ref{S:matrix}.)


\subsection{Reduction to Hamiltonian}

As noted earlier (\ref{E:pm}), on reducing F $£$ L $F_\A$ branches into selfdual and anti-selfdual halves, while under F $£$ G it splits into its $ $ and $§$ halves; one division is the sum and difference of the other.  
\begin{align}
{\rm F} £ {\rm G} & : F = (F_ ,F_§) \\
{\rm F} £ {\rm L} & : F = (F^{(+)},F^{(-)}) = ú^0 F_  à F_§
\end{align}
A similar doubling occurs for $»_\A$ because of its chirality after reduction, which for lower D was eliminated by selfduality using a 4-index Levi-Civita tensor, but is now resolved using part of the $\V$ constraint.

Furthermore, we can directly identify $F_ $ with $P$ of the $\H$ formalism, so the reduction from $\L$ to $\H$ formalism can be made directly in terms of F $£$ G.
For example, we can represent the branching for
\begin{equation}
\bf
56 £ 1_3 + 1_{-3} + 27_1 + 27'_{-1}¼,â133 £ 1_0 + 27_{-2} + 27'_2 + 78_0
\end{equation}
(where the subscripts indicate the GL(1) of E$_{6(+6)}ð$ GL(1) $¤$ E$_{7(7)}$) in matrix notation as
\begin{equation}
F^\A = \bordermatrix{ & \cr 1_3 & F \cr 1_{-3} & F' \cr 27_1 & F^a \cr 27'_{-1} & F_a \cr}¼,â
X^{\A\B} = \bordermatrix{ & 1_3 & 1_{-3} & 27_1 & 27'_{-1} \cr
			1_3 & 0 & 3X & 0 & X_b \cr
			1_{-3} & 3X & 0 & X^b & 0 \cr
			27_1 & 0 & X^a & d^{abc}X_c & \ff^A{}_b{}^a X_A + ¶_b^a X \cr
			27'_{-1} & X_a & 0 & \ff^A{}_a{}^b X_A + ¶_b^a X & d_{abc}X^c \cr}
\end{equation}
where $X^a,X_a,$ and $X^A$ are the {\bf 27}, {\bf 27$'$}, and {\bf 78} (adjoint) of E$_6$.
(Coefficients that are not merely conventional, for the E$_7$ adjoint representation, are determined by the commutation relations in Appendix \ref{S:norm}.
$\ff^A{}_a{}^b$ are the {\bf 27} representation of the E$_6$ generators.)
Using also the Sp(56) metric
\begin{equation}
C^{\A\B} = \bordermatrix{ & 1_3 & 1_{-3} & 27_1 & 27'_{-1} \cr
			1_3 & 0 & 1 & 0 & 0 \cr
			1_{-3} & -1 & 0 & 0 & 0 \cr
			27_1 & 0 & 0 & 0 & ¶_b{}^a \cr
			27'_{-1} & 0 & 0 & -¶_a{}^b & 0 \cr}¼,â
C^{\A\C}C_{\B\C} = ¶^\A_\B
\end{equation}
to raise/lower indices (e.g., $»_\A=»^\B C_{\B\A}$), we can then decompose the equation
\begin{equation}
F^\A = »_\B X^{\B\A}âÜ
\end{equation}
\vskip-.4in
\begin{subequations}
\begin{align}
F & = 3»X + »^b X_b \\
F' & = - 3»'X - »_b X^b \\
F^a & = »X^a - d^{abc}»_b X_c + »^b \ff^A{}_b{}^a X_A+ »^a X\\
F_a & = - »' X_a - »_b \ff^A{}_a{}^bX_A + d_{abc}»^b X^c - »_a X
\end{align}
\end{subequations}

Similarly we can find the components of $\V$, which is the same representation as $X$, by considering $X^{\A\B}»_\A »_\B=X^{\A\B}\V_{\A\B}$, and separating the coefficients of the components of $X$:
\begin{subequations}
\begin{align}
\V & = 3»»' + »_a »^a \\
\V^a & = -2»'»^a + d^{abc}»_b »_c \\
\V_a & = -2»»_a + d_{abc}»^b »^c \\
\V_A & = \ff_{Ab}{}^a »_a »^b
\end{align}
\end{subequations}
We can partially solve $\V^{\A\B}=0$ as
\begin{equation}
»' = »_a = 0¼,â\hbox{leavingâ}\V_a =d_{abc}»^b »^c = 0
\end{equation}
(Or we might impose selfduality for $»$, yielding the same result, in addition to $\V$.  See the discussion of the identity (\ref{E:selfd}).)
Then we identify
\begin{equation}
»^a = »_§^a¼,â» = »_ 
\end{equation}

The surviving terms in $F_A$ are then
\begin{equation}
F^{(à)}_a = P_a à d_{abc}»^b X^c¼,âF^{(à)} = P à 0
\end{equation}
where
\begin{equation}
ú^{ab}{}_0 P_b = ÀX{}^a + »^b \ff_{Ab}{}^aX^A + »^a X¼,âP = ÀX + »^b X_b
\end{equation}
Here $X^a$ appears as the usual ``physical" part of $X^{\A\B}$, while $X_a$ and $X^A$ are the usual Lagrange multipliers for Gauss's law.

On the other hand, the ``scalar" $X$ is something new:  It appears as physical, and so not a gauge field, in $P$ (with a $»_ $), and so can't be a gauge field in $P_a$.  
(The complete gauge transformations can be found in Appendix \ref{S:para}, as derived from (\ref{E:para}).)
However, the corresponding selfduality condition is $F^{(-)}=P-0=0$, so it can be eliminated.  Similarly, its apparent ``Gauss law" $»^aP_a=0$ is already implied by the other selfduality condition, using the $\V$ constraint.  What's left then agrees with the results of \cite{Linch:2016ipx} for D = 5.  


\subsection{F \texorpdfstring{$\rightarrow$}{\textrightarrow} M}

Solving the $\S$ constraint $d^{abc}F_bF_c=0$ gives us the reduction F $£$ M. Following the decomposition of E$_6 £$ SL(6) $\otimes$ SL(2), we have:
\begin{equation}
\bf
27 £ (15,1)+(6',2),â27' £ (15',1)+(6,2),â78 £ (35,1)+(20,2)+(1,3)
\end{equation}
Now using ``$a$" for an SL(6) ``vector" index and ``$i$" for SL(2), the section condition is then split into:
\begin{subequations}
\begin{align}
P_{ab}p^b_i+p_{ab}P^b_i&=0\\
\epsilon^{abcdef}P_{cd}p_{ef}+\epsilon^{ij}P^a_ip^b_j&=0
\end{align}
\end{subequations}
and a copy with the operators $P_{ab}$,$P_i^a$ being replaced by their zero mode.
We can pick a gauge in which only $P^a_1$  survives.
Since the worldvolume index lives in the representation $\bf 27'$, the world volume section constraint splits in a similar way.
This gives us the field strengths
\begin{subequations}
\begin{align}
F_{ab}=&-\epsilon_{abcdef}\partial^{e2}X^{f1}\\
F_{a1}=&P_{a1}\\
F_{a2}=&\partial_{ab}X^{b1}
\end{align}
\end{subequations}
The Gauss constraint now looks like:
\begin{subequations}
\begin{align}
\partial^{a1}P_{b1}&=0\\
\partial^{a1}P_{a1}&=0\\
\epsilon^{abcdef}\partial_{de}P_{f1}&=0
\end{align}
\end{subequations}


\subsection{F \texorpdfstring{$\rightarrow$}{\textrightarrow} T}

Solving the Gauss constraint $(\partial _a P^b)_{78}=0$ and the remaining worldvolume section condition $d_{abc}\partial ^b \partial ^c =0$ explicitly breaks the $E_6$ group to O(5,5), and as a result we go back to T-theory in 5D.

Decomposing E$_6 £$ O(5,5), we have:   
\begin{equation}
\bf
27 £ 10_2 + 16_{-1} + 1_{-4},â27' £ 10_{-2} + 16'_{1} + 1_4,â78 £ 45_{0} + 16'_{-3}+16_3+1_0
\end{equation}

Therefore, in terms of O(5,5) group notation (scalar $s$, spinor $\alpha$ , vector $a$, and $\eta_{ab}$,$C_{\alpha\beta}$ for contraction if needed), we can rewrite our constraints as:
\begin{subequations}
\begin{align}
2\partial _a \partial _s +\gamma _{a\alpha\beta}\partial ^{\alpha}\partial^{\beta}&=0\\
\partial _a\partial ^a &=0\\
\gamma_{\beta\alpha a}\partial ^a \partial ^\alpha &=0
\end{align}
\end{subequations}
\begin{subequations}
\begin{align}
\partial _s P_s +\partial ^a P_a +\partial ^{\alpha}P_{\alpha}&=0\\
\partial_s P_{\alpha}+\gamma_{a\alpha\beta}\partial^{\beta}P^{\alpha}&=0\\
\partial^\alpha P_s+\gamma_{a\beta}^\alpha \partial ^a P^\beta &=0\\
\partial_{[a}P_{b]}+\gamma_{ab}^{\alpha\beta}\partial_\alpha P_\beta &=0
\end{align}
\end{subequations}
It is easy to check that picking a gauge with $\partial _\alpha ,\partial_a = 0$ solves the first set of equations. Applying them on the second set tells us that only the vectors $P_a$ survive.
Then the remaining currents (selfdual field strengths) are:
\begin{equation}
F_a^{(+)}=P_a +\eta_{ab}\partial_s X^{b}
\end{equation} 
and the remaining Virasoro constraint is:
\begin{equation}
d^{sab}F_a^{(+)}F_b^{(+)}= \eta^{ab}F_a^{(+)}F_b^{(+)}=0
\end{equation}
which is the standard algebra for T-theory.


\section{Conclusions}

We plan to consider the coupling of massless background fields in the Lagrangian formalism in a future paper; previously (even in T-theory) it was possible to manifest all the symmetries of the background only in the Hamiltonian approach.
There is the related question of how this coupling disentangles itself from that of the worldvolume metric, which had similar problems in the Lagrangian approach:  Unlike S-theory, the two kinds of metric carry related indices.

These and other questions lead us to consider ``zeroth-quantization" in terms of a space that bears the same relation to the worldvolume that the worldvolume does to spacetime.  
(Early papers on zeroth-quantization of string theory \cite{Green:1987cw,Kutasov:1996fp,Kutasov:1996zm,Kutasov:1996vh} also involved selfdual theories \cite{Ooguri:1990ww,Ooguri:1991fp,Marcus:1992xt,Siegel:1992ev}.)
There are a number of reasons why this might be expected or preferable as a starting point for formulating F-theory:

\bitem{(1)}  There is the analogy of $\V$ (worldvolume) sectioning to $\S$ (spacetime) sectioning.  
As the onset of $\V$ constraints is with higher dimensions of F-theory, while the other constraints start from D = 1, there is the suggestion that the constraints might be simpler in zeroth-quantization than in first.
The ``metric" for the $\S$ constraint is dual to that appearing with the central charge term in the current algebra, suggesting a similar role for $\V$ in defining a zeroth-quantized current algebra.

\bitem{(2)}  The introduction of zeroth-quantized ghosts \cite{Siegel:2016dek} 
(fermionic partners for $§$) may be necessary to quantize in a way that higher symmetries are preserved in quantum calculations (except for unitary gauges for external polarizations in S-matrices).

\bitem{(3)}  G-symmetry apparently becomes infinite-dimensional for D $>$ 7, and F-symmetry for D¼$>$¼5.  But E$_9$ is recognized as a 2D current algebra of the finite-dimensional group E$_8$, corresponding to a zeroth-quantized worldsheet.  Similar constructions may be possible for the other infinite-dimensional algebras.

\noindent An interesting question is then whether in some sense the zeroth-quantized theory can be considered Type II in the same sense as the first-quantized.  To this end we consider a construction of F-theory as the direct product of ``left" and ``right" open-string theories.  For example, for D = 5 the E$_7$/SU*(8) structure is represented by the coset {\bf 133} $-$ {\bf 63} = {\bf 70}, a ``4-form" of SU*(8) (as for the scalars of 4D N = 8 supergravity), while $»$ and $F$ are each {\bf 56} = {\bf 28} + {\bf 28$'$}, dual ``2-forms" of SU*(8) (as for the vectors of 4D N = 8 supergravity).  If we halve the size of the forms (to get just left or just right), and correspondingly also reduce SU*(8) to SU*(4) (to preserve Hodge duality of forms), we also find E$_7$ reduced to SO(6,1), as SO(6,1)/SU*(4) = AdS$_6$ = {\bf 21} $-$ {\bf 15} = {\bf 6} (2-forms, as for the scalars of 4D N = 4 Yang-Mills), while $»=F=$ {\bf 8} = {\bf 4} + {\bf 4$'$} (dual 1-forms, as for the spinors).


\section*{Acknowledgements}

WS thanks William Linch for help with early research on this paper.
WS is supported by NSF grant PHY-1620628.


\appendix

\section{Matrix algebra}
\label{S:matrix}

Much of the GL (or U or U*) group theory can be done more conveniently by matrix methods than by manipulation of indices on Kronecker $¶$'s.
For purposes of the next few paragraphs we'll treat $\un A$ as the index for a general matrix, while $S$ is that for a symmetric matrix, and $A$ antisymmetric, so
\begin{equation}
\un A= (S,A)
\end{equation}
Using elements $M$ of a matrix basis we write, in terms of defining-representation (``spinor") indices $Œ$,
\begin{equation}
M_{\un A} = M^{\un A} = M_Œ{}^º¼,âM_S = M_{(Œº)}¼,âM_A = M_{[Œº]}¼,âM^S = M^{(Œº)}¼,âM^A = M^{[Œº]}
\end{equation}
where the (anti)symmetrization of indices is used only as a reminder of the symmetry and
\begin{equation}
(M_{\un A})^{\un B} ­ ¶_{\un A}^{\un B}
\end{equation}
where ``$\un A$" labels which matrix and ``$\un B$" labels which component of that matrix, necessarily also decomposed into spinor indices to make it a matrix and not a vector.

Using ``$Ò¼Ô$" to indicate the supertrace we have, for arbitrary matrices $N,P,Q,R$, identities such as
\begin{equation}
ÒNM^{\un A}ÔÒM_{\un A}PÔ = ÒNPÔ
\end{equation}
\begin{equation}
ÒNM^SÔÒM_S PÔ = üÒN(P+P^T)Ô¼,âÒNM^AÔÒM_A PÔ = üÒN(P-P^T)Ô
\end{equation}
\begin{equation}
ÒNPM^{\un A}ÔÒM_{\un A}QRÔ = ÒNPQRÔ = \hbox{cyclic} = ÒPQM^{\un A}ÔÒM_{\un A}RNÔ
\end{equation}

As an example, we give a derivation of the identity (\ref{E:beta})
$$
\ff^{D(\A\B} º_D{}^{\C)\rm E} = 0
$$
where $\ff$ is symmetric in its two {\bf 56} indices because (see, e.g., \cite{Yamatsu:2015npn})
$$
(56°56)_S = 133¢1463¼,â(56°56)_A = 1¢1539
$$
We do this by breaking E$_{7(7)}£$ U*(8), proving it for one of the resulting equations using U*(8) matrix methods, then concluding that the full identity applies by E$_{7(7)}$ symmetry.
(We will assume $(\A\B\C)\rm E$ is irreducible in E$_7$.)

Since
$$
56 £ 28+28'¼,â133 £ 70+63¼,â912 £ 36+36'+420+420'
$$
we choose the representations that are 8$ð$8 matrices of various symmetries,
$$ \A,\B £ 28¼,â\C £ 28'¼,âD £ 63¼,â{\rm E} £ 36' $$
(Here the dual indicated by the prime relates covariant to contravariant indices.  The SU*(8) adjoint {\bf 63} has one index up and one down, while the ``4-form" {\bf 70} is ``Hodge dual" between its covariant and contravariant versions.  The {\bf 420} is mixed symmetry, with 3-form indices down and 1 index up, traceless.)

The trace on the index $D$ can be taken as on a general matrix {\bf 1} + {\bf 63}, since the singlet piece won't contribute to $º$.
Then $\ff$ has only ({\bf 28})({\bf 28}$'$) and ({\bf 28}$'$)({\bf 28}) pieces (to always contract up indices with down), which are transposes (without sign, by the full $\ff$'s symmetry), while $º$ has only ({\bf 28})({\bf 36}$'$).  This case of the identity to be proven then reduces to simply
$$
ÒM_\A M^\C M_\B M^{\rm E} + M_\B M^\C M_\A M^{\rm E}Ô = 0
$$
which follows from relating the former term to the latter by transposition.  (The {\bf 28}'s are antisymmetric, while the {\bf 36} is symmetric.)


\section{Normalization factors}
\label{S:norm}

The weight factors appearing in the group decomposization can be determined by comparing the closure of our ansatz. Since there is an overall arbitrariness when defining the variables, we can normalize our components as
\begin{equation}
X_\A{}^\B =
\begin{pmatrix}
\alpha X& 0 & X^b &0 \\
0& -\alpha X&0&-X_b\\
X_a&0&\ff_{Aa}{}^bX^A+\delta_a^bX&\gamma d_{abc}X^c\\
0&-X^a&\beta d^{abc}X_c&-\ff_{Ab}{}^aX^A-\delta_b^aX
\end{pmatrix}
\end{equation}
Then we can compute the commutator $[X,Y]=Z$, and write the expression for $Z$:
\begin{subequations}
\begin{align}
\alpha Z&=X^aY_a-X\leftrightarrow Y\\
Z^a&=\alpha YX^a+\ff_{Ab}{}^aX^AY^b+XY^a-X\leftrightarrow Y\\
d_{adc}Z^c&=(\ff_{Aa}{}^bX^A+\delta_a^bX)d_{bcd}Y^c-(\ff_{Ad}{}^cY^A+\delta_d^cY)d_{abc}X^b-X\leftrightarrow Y\\
\ff_{Aa}{}^bZ^A+\delta_a^bZ&=X_aY^b+(\ff_{Aa}{}^cX^A+\delta_a^cX)(\ff_{Bc}{}^bY^B+\delta_c^bY)+\beta\gamma d_{adc}d^{bde}X^cY_e-X\leftrightarrow Y
\end{align}
\end{subequations}
The result can be read off: $\alpha=3$, $\beta=\gamma=1$.


In evaluating this commutator we have used the projection operator decomposition of {\bf 27} $°$ {\bf 27$'$}.
The fundamental identity required for D = 5 bosonic F-theory in $\H$ E$_6$ notation
is the Springer relation
\begin{equation}
d^{efg}(d_{eab}d_{cdf} + d_{eac}d_{dbf} + d_{ead}d_{bcf}) = ¶_a^g d_{bcd} + ¶_b^g d_{cda} + ¶_c^g d_{dab} + ¶_d^g d_{abc} .
\end{equation}
where $d_{abc}$ is the totally symmetric invariant tensor of E$_6$ in terms of indices ``$a$" for the {\bf 27} and {\bf 27$'$}.
This normalization of $d^{abc}$ vs.¼its dual $d_{abc}$ corresponds to
\begin{equation}
d^{acd}d_{bcd} = 10 ¶_b^a
\end{equation}
found from the above by tracing.  
At this point we convert to matrix notation, treating {\bf 27 $°$ 27$'$} as a vector.  Then
\begin{equation}
Y_a{}^b{}_d{}^c ­ d_{ade}d^{bce}
\end{equation}
acts as a matrix in this space
We also introduce the matrix that picks out the trace piece,
\begin{equation}
T_a{}^b{}_d{}^c ­ \f1{27}¶_a^b ¶_d^c
\end{equation}
Contracting a $d$ with the Springer relation, we then have
\begin{equation}
Y^2 +4Y -5I = 135T
\end{equation}
This identity tells us how to separate the {\bf 1 + 78 + 650} pieces of {\bf 27 $°$ 27$'$}:  The decomposition into projection operators is (solving for orthonomality)
\begin{equation}
I = T + \f16 (I+9T-Y) + \f16 (5I-15T+Y)
\end{equation}
(We also used $TY=YT=10T$.
The trace of each projector gives the dimension of its representation.)
In particular, we have for the ubiquitous matrix $U$
\begin{equation}
U ­ I - Y = -9[T] + 6[\f16(I+9T-Y)]
\end{equation}
Introducing {\bf 78 $¢$ 1} indices ``$\un A$", where $\un A = (A,0)$,
we choose a normalization such that
\begin{equation}
d_{ade}d^{bce} = ¶_a^c ¶_d^b - ¶_a^b ¶_d^c + \f43\ff_{\un Ea}{}^b \ff^{\un E}{}_d{}^c¼,â\ff_{0a}{}^b = \ff^0{}_a{}^b = ¶_a^b
\end{equation}
where $\ff$ are the generators of E$_6 ð$GL(1) in the {\bf 27} representation, since $\un A$ is the adjoint.
({\bf 27} is dual to {\bf 27$'$}, while {\bf 78} is selfdual, so the metric $ú_{\un A\un B}$ and its inverse exist.)


\section{Gauge parameters}
\label{S:para}

Breaking the $\bf  912$ of $E_7$ into representations of $E_6$, we have:
\begin{equation}
\bf
912 £ 351'_1+351_{-1}+27_{-1}+27'_{1}+78_3+78_{-3}
\end{equation} 
with $\bf 351'=(27°27)_A$ (and similar for $\bf 351$ in terms of $\bf 27'$).
Then we can decompose the gauge transformation into:
\begin{subequations}
\begin{align}
\delta X_A&= \ff_{Ac}{}^d d^{abc}\partial_b\Lambda_{ad}+\ff_{Ad}{}^c d_{abc}\partial^b\Lambda^{ad}+\ff_{Ac}{}^d\partial_d\Lambda^c+\ff_{Ac}{}^d\partial^c\Lambda_d-\partial'\Lambda_{1A}-À\Lambda_{2A}\\
\delta X_a &=\partial ^b\Lambda_{ab}+d_{abc}\partial^b\Lambda^c-3À\Lambda_a+\ff_{Aa}{}^b\partial_b \Lambda_{1A}\\
\delta X^a&= \partial _b\Lambda^{ab}+d^{abc}\partial_b\Lambda_c-3\partial'\Lambda^a+\ff_{Ab}{}^a\partial^b \Lambda_{2A} \\
\delta X&=\partial_a \Lambda^a +\partial^a \Lambda_a
\end{align}
\end{subequations}
After applying the solution of the worldvolume section condition, we are left with:
\begin{subequations}
\begin{align}
\delta X_A&=\ff_{Ad}{}^c d_{abc}\partial^b\Lambda^{ad}+\ff_{Ac}{}^d\partial^c\Lambda_d-À\Lambda_{2A}\\
\delta X_a &=\partial ^b\Lambda_{ab}+d_{abc}\partial^b\Lambda^c-3À\Lambda_a\\
\delta X^a&= \ff_{Ab}{}^a\partial^b \Lambda_{2A}\\
\delta X&=\partial^a \Lambda_a
\end{align}
\end{subequations}
As a check, we can evaluate the gauge transformation of the field strengths and indeed they are invariant, up to the remaining worldvolume section $d_{abc}\partial^b\partial^c=0$.


\small
\linespread{1.1}\selectfont
\raggedright

\bibliography{Bibtex}
\bibliographystyle{utphys}


\end{document}